\documentstyle[11pt]{article}
\begin{document}
\normalsize
\baselineskip .1in
\flushbottom
\parindent 0.25in
\oddsidemargin 0.75in         
\evensidemargin 0.75in
\topmargin=1in           
\headheight=0.1in           
\headsep= 0.2in           
\footskip=0.3in           
\footheight=0.3in        
\textheight = 6.9in
\textwidth 5.1in         

\def\a{\alpha}
\def\b{\beta}
\def\ch{\chi}
\def\d{\delta}
\def\e{\epsilon}
\def\f{\phi}
\def\g{\gamma}
\def\h{\eta}
\def\i{\iota}
\def\j{\psi}
\def\k{\kappa}
\def\l{\lambda}
\def\m{\mu}
\def\n{\nu}
\def\o{\omega}
\def\p{\pi}
\def\q{\theta}
\def\r{\rho}
\def\s{\sigma}
\def\t{\tau}
\def\u{\upsilon}
\def\x{\xi}
\def\z{\zeta}
\def\D{\Delta}
\def\F{\Phi}
\def\G{\Gamma}
\def\J{\Psi}
\def\L{\Lambda}
\def\O{\Omega}
\def\P{\Pi}
\def\S{\Sigma}
\def\U{\Upsilon}
\def\X{\Xi} 
\def\T{\Theta}
\def\vf{\varphi}

\def\Ab{\bar{A}}
\def\gi{g^{-1}}
\def\li{{ 1 \over \l } }
\def\lb{\l^{*}}
\def\zb{\bar{z}}
\def\ub{u^{*}}
\def\vb{v^{*}}
\def\Tb{\bar{T}}
\def\pp {\partial }
\def\pb {\bar{\partial }}
\def\be{\begin{equation}}
\def\ee{\end{equation}}
\def\ben{\begin{eqnarray}}
\def\een{\end{eqnarray}}

\begin{center}
{\bf Nonabelian sine-Gordon theory
 and its application to nonlinear optics }
\vskip 1em
 Q-Han Park, H.J. Shin
 \vskip 1em
{\it \footnotesize  Department of Physics and 
Research Institute of Basic Sciences,
Kyunghee University,
Seoul, 130-701, Korea}
\end{center}
\vskip 1em
\begin{minipage}{5.0truein}
                 \footnotesize
                 \parindent=0pt 

Using a field theory generalization of the 
spinning top motion, we construct nonabelian generalizations of the 
sine-Gordon theory according to each symmetric spaces. 
A Lagrangian formulation of these generalized sine-Gordon theories 
is given in terms of a deformed gauged Wess-Zumino-Witten action 
which also accounts for integrably perturbed coset conformal field 
theories. As for physical applications, we show that they become 
precisely the effective field theories of self-induced transparency 
in nonlinear optics. This provides a dictionary between field theory 
and nonlinear optics.
                 \par
                 \end{minipage}
                 \vskip 2em 

Among many integrable equations, the sine-Gordon equation is one of 
the most well-known equation which finds countless applications in a 
wide range of physics due to its ubiquitous nature. In many cases, 
however, it is desirable to have a ``generalized" sine-Gordon equation 
in order to accomodate more realistic physical systems. In this talk, 
I will show that such generalizations are indeed possible. They are 
constructed according to each coset $G/H$ and shown to provide 
a Lagrangian formulation of integrably perturbed coset conformal field 
theories. Surprisingly, when $G/H$ is restricted to a Hermitian 
symmetric space, the generalized sine-Gordon equation finds unexpected 
applications in nonlinear optics. In particular, when $G/H = SU(2)/U(1)$, 
it describes the two-level self-induced transparency (SIT),
a phenomenon of anomalously low energy loss in coherent optical pulse 
propagation\cite{McCall}.

In order to gain  physical insight as well as a handle for the 
generalization, we first introduce the SIT equation which in 
the sharp line limit is given by the Maxwell equation under the ``slowly 
varying envelope approximation"
\be
\pb E + 2 \b P = 0 
\label{max}
\ee
and the optical Bloch equation
\ben
\pp D - E^{*}P - EP^{*} &=& 0 \nonumber \\
\pp P + 2i\xi P + 2ED &=& 0 
\label{bloch}
\een
where $\xi  = w-w_{0} \ , \ \pp \equiv \pp /\pp z \ , \ \pb \equiv \pp / 
\pp \bar{z} ~, z= t-x/c , \bar{z} = x/c$ and $E,P$ and $ D$ represent the 
electric field, the polarization and the population inversion 
respectively. Here, we will not concern about the details of nonlinear 
optics and for the sake of this talk, it suffices to say that 
Eqs.(\ref{max}) and (\ref{bloch}) are coupled nonlinear partial 
differential equations in 1+1-dimensional spacetime with two complex 
fields $E$ and $P$ and one real field $D$ together with constant 
parameters $\b$ and $\xi $. Note that the optical Bloch equation 
admits an interpretation as a spinning top equation like the 
corresponding magnetic resonance equations \cite{bloch}. To see this, 
denote real and imaginary parts of $E$ and $P$ by $E = E_{R} + 
iE_{I}, ~ P = P_{R} + iP_{I}$. Then, the Bloch equation (\ref{bloch}) 
can be expressed as 
\be
\pp \vec{S} = \vec{\Omega } \times \vec{S}
\label{topeq}
\ee
that is, it describes a spinning top motion where the electric dipole 
``pseudospin" vector $\vec{S} = (P_{R}, ~ P_{I}, ~ D)$  precesses 
about the ``torque" vector $\vec{\Omega } = (2E_{I}, ~ -2E_{R}, ~-2\xi )$. 
This clearly shows that the length of the vector $\vec{S}$ is conserved,
\be
|\vec{S}|^{2} = P_{R}^{2} + P_{I}^{2} + D^{2} = 1,
\label{prob}
\ee
where the length equals unity due to the conservation of probability.
In case $P_{I} = 0$, we may solve Eq.(\ref{prob}) for 
$P_{R} = -\sin{2\vf } $ and $D = \cos{2\vf }$. Then, Eq.(\ref{topeq}) 
can be solved easily by taking $E = \pp \vf $ which changes the Maxwell 
equation to the sine-Gordon equation. The Maxwell equation determines the 
strength of the torque vector along the $\zb $-axis which 
agrees with the conventional mechanical interpretation of the sine-Gordon 
equation as a continuum limit of the infinite chain of coupled pendulum 
equations, i.e. it becomes a field theory generalization of a pendulum. 
This shows that the SIT equation, without the assumption $P_{I} = 0$, 
is a field theory generalization of a spinning top which includes the 
sine-Gordon theory as a special case.
The key observation leading to the nonabelianization of the sine-Gordon 
equation is that the constraint Eq.(\ref{prob}) can be solved generally 
in terms of an $SU(2)$ matrix potential variable $g$ by
\be
\pmatrix{D & P \cr P^{*} & -D} = g^{-1}\s_{3}g, ~~~ \s_{3} = 
\pmatrix{1 & 0 \cr 0 & -1} .
\label{gspin}
\ee
Then, the Bloch equation arises from the identity,
\be
\pp (g^{-1} \s_{3} g) = [ g^{-1}\s_{3}g, ~ g^{-1} \pp g],
\label{identity}
\ee
if we take
\be
g^{-1} \pp g -R =  \pmatrix{ i\xi  & -E \cr E^{*} & -i\xi } ,
\label{idelec}
\ee 
where $R$ is an arbitrary matrix commuting with $g^{-1}\s_{3}g$ which 
will be determined later. Finally, the Maxwell equation becomes
\ben
\pb (g^{-1} \pp g -R) &=& \pmatrix{ 0 & -\pb E \cr \pb E^{*} & 0 } 
= - \left[ \  \pmatrix{ i\b  & 0 \cr 0 & -i\b  } \ , 
~  i \pmatrix{D & P \cr P^{*} & -D }
\right] \nonumber \\ 
&=&  \b[ \s_{3}, ~ g^{-1}\s_{3} g].
\label{maxpot}
\een
Thus, the SIT equation changes into a single nonlinear sigma model-type 
equation up to an undetermined quantity $R$. 

Now, we show that $R$ is fixed when we consider a Largrangian formulation 
of the SIT equation. In fact, we will construct a Lagrangian which is more 
general than the SIT case in terms of the gauged Wess-Zumino-Witten action 
as follows;
\be
S = S_{WZW}+ S_{\mbox{gauge}}  - S_{\mbox{pot}} 
\label{action} 
\ee
where $S_{WZW}$ is the usual group $G$ Wess-Zumino-Witten action 
and $S_{\mbox{gauge}} $ is the gauging part,
\be
S_{\mbox{gauge}} = 
{1 \over 2\pi }\int \mbox {Tr} (- A\pb g \gi + \Ab \gi \pp g
 + Ag\Ab \gi - A\Ab ),
\ee
which gauges the anomaly free vector subgroup $H \in G$.
This gauged Wess-Zumino-Witten action $ S_{WZW} + S_{\mbox{gauge}}  $ 
has been identified as an action of $G/H$ coset conformal field theories
\cite{coset}. The potential term $S_{\mbox{pot}}$ is added in such 
a way that the integrability of the model is preserved while the conformal 
symmetry is broken. A general construction of $S_{\mbox{pot}}$ is given by 
a triplet of Lie groups $F \supset G \supset H$ for every symmetric space 
$F/G$, where the Lie algebra decomposition ${\bf f} = {\bf g} 
\oplus {\bf k} $ satisfies the commutation relations,
\be
[{\bf g} ~ , ~ {\bf g}] \subset {\bf g} ~ , ~ [ {\bf g} ~ , ~ {\bf k}] 
\subset {\bf k} ~ , ~ [{\bf k} ~ , ~ {\bf k} ] \subset {\bf g} ~ .
\label{algebra}
\ee
We take $T$ and $\Tb$ as elements of $ {\bf k}$ and  define  
$ {\bf h} $ as the simultaneous centralizer of $T$ and $\Tb $, i.e. 
${\bf h} = C_{{\bf g}}(T, \Tb ) = \{ B \in {\bf g} \ : \ [B ~ , ~ T] = 
0 = [B ~ , ~ \Tb]\} $ with $H$ its associated Lie group. 
With these specifications, the potential $S_{\mbox{pot}}$ is given by 
\be
S_{\mbox{pot}}= {\b \over 2\pi }\int \mbox{Tr}gT \gi \Tb .
\label{potential}
\ee
The (classical) integrability can be demonstrated by expressing the 
equation of motion arising from the action (\ref{action}) in a zero 
curvature form with a spectral parameter $\l $,
\be
\d_{g}S  =0 \leftrightarrow 
[\  \pp + \gi \pp g + \gi A g + \b\l T \ , \ \pb + \Ab + 
{1 \over \l }\gi \Tb g \ ] = 0
\label{zeroeqn}
\ee
Also, due to the absence of the kinetic terms, $A, \Ab $ act as Lagrange 
multipliers which result in the constraint equations;
\be
( \ - \pb g \gi + g\Ab \gi - \Ab \  )_{\bf h} 
= 0  = ( \  \gi \pp g  +\gi A g - A \ )_{\bf h}
\label{constraint}
\ee
where the subscript ${\bf h}$ denotes the projection to the 
subalgebra ${\bf h}$.
Explicit expressions of equations are given in Ref.\cite{bps} for various 
cases of symmetric spaces, especially for the type I symmetric spaces: 
$F/G = SO(n+1)/SO(n), ~ SU(n)/SO(n),   ~ SU(n+1)/U(n), ~ Sp(n)/U(n)$. These 
equations include the sine-Gordon equation as a special case thereby called 
as symmetric space sine-Gordon (SSSG) equations \cite{bps}. 

A couple of interesting cases of SSSG equations are in order in terms of a 
triplet of symmetric spaces ($F,~ G, ~ H)$),
\vskip .1in
$\underline{ (F, ~ G, ~ H) = 
(SU(4), ~ SU(2)\times SU(2) \times U(1), ~ SU(2))} $
\vskip .1in
The $U(1)$ factor can be decoupled consistently and the model describes 
an integrable perturbation of the minimal model in 
conformal field theory which itself can be defined by the $SU(2) \times 
SU(2)/SU(2)$-gauged WZW model \cite{shin2}. If we choose 
\be
g = \pmatrix{g_{1} \in SU(2) & 0 \cr 0 & g_{2} \in SU(2) }
, ~ T = \Tb = \pmatrix{0 & i {\bf 1}_{2 \times 2} \cr i {\bf 1}_{2 \times 2} 
& 0 },
\ee
then, the model describes the integrable deformation of the minimal 
conformal theory for the critical Ising model by the operator 
$\Phi_{(2,1)}$. 
This case has been called as a matrix sine-Gordon theory and its classical 
behavior and soliton solutions have been analyzed in \cite{shin1}. 
Another interesting deformation arises with a choice
\be
g = g_{1}g_{2} \in SU(2) \otimes SU(2), ~ T = \Tb = \sum_{a = 1}^{3}L^{a} 
\otimes M^{a} 
\ee
for $ L^{a} (M^{a}) $ generators of $ su(2)$ and this describes the 
deformation by the operator $\Phi_{(3,1)}$ \cite{shin2}.
\vskip .1in
\underline{ $(F,~ G, ~H)= (SO(5),~ SO(3) \times U(1),~ SO(2))$}
\vskip .1in
With the $U(1)$ decoupled, this case is known as the 
complex sine-Gordon equation which in turn accounts for the SIT equation 
(\ref{max}) and (\ref{bloch}) as explained below. These two cases of SSSG 
are special cases of the nonabelian Toda theory with $N=1$ 
grading \cite{hollowood}.

In case of compact symmetric spaces of type II, e.g. 
symmetric spaces of the form $G \times G/G$, the elements $g$ and 
$T$ take the form $g \otimes g$ and $T \otimes 1 - 1 \otimes T$ 
(and similarly for $\Tb $). In which case, the model becomes 
effectively equivalent to the case where $T, ~ \Tb$ belong to the 
Lie algebra ${\bf g}$. Thus the model is specified by the coset $G/H$ 
where $H$ is the stability subgroup of $T$ and $\Tb $ for $T, ~ \Tb 
\in {\bf g }$. In particular, when $G/H$ is further restricted to 
Hermitian symmetric space, a symmetric space equipped Hermitian structure,  
the SSSG equation finds a nice physical application, i.e. it becomes 
precisely the SIT equations for various atomic systems. 
In this case, the adjoint action of $T$ defines a complex structure 
on $G/H$. For example, in the simplest $CP^{1}$ case where 
$G/H = SU(2)/U(1) \approx CP^{1}$ , we may choose $T=\Tb = i\s_{3}$ 
and fix the vector gauge invariance of the action (\ref{action}) by 
\be
A = i\x \s_{3} \ , \ \Ab =0
\ee
for a constant $\x $. Such a gauge fixing is possible due to the flatness 
of $A, \Ab$. Also, we parameterize the $2\times 2$ matrix $g$ by
\be
g=e^{i\eta \s_{3}}e^{i\varphi (\cos{\q }\s_{1} -\sin{\q }\s_{2})}e^{i\eta 
\s_{3}}= \pmatrix{ e^{2i\eta }\cos{\varphi } & i\sin{\varphi }e^{i\q } \cr
i\sin{\varphi }e^{-i\q } & e^{-2i\eta }\cos{\varphi } }\ .
\label{param}
\ee
and identify $E, P$ and $D$ with $g$ through the relation
\be
g^{-1}\pp g + \x g^{-1}Tg - \x T = \pmatrix{ 0 & -E  \cr E^{*} & 0 } \ \ 
, \ \ 
g^{-1}\bar{T}g = -i \pmatrix{ D & P \cr P^{*} & -D }
\ee
where we have imposed the constraint (\ref{constraint}). Then, it is a 
straightforward exercise to show that the zero curvature equation 
(\ref{zeroeqn}) reduces to the SIT equation.

Other cases of Hermitian symmetric spaces are also associated with various 
multi-level SIT systems with resonant transitions and 
many new aspects of SIT arising from these identifications and 
their physical implications are explained in [8].

\end{document}